# Recent Experimental Progress of Fractional Quantum Hall Effect: 5/2 Filling State and Graphene

X. Lin, R. R. Du and X. C. Xie

*International Center for Quantum Materials, Peking University, Beijing, People's Republic of China 100871*

**ABSTRACT**

The phenomenon of fractional quantum Hall effect (FQHE) was first experimentally observed 33 years ago. FQHE involves strong Coulomb interactions and correlations among the electrons, which leads to quasiparticles with fractional elementary charge. Three decades later, the field of FQHE is still active with new discoveries and new technical developments. A significant portion of attention in FQHE has been dedicated to filling factor 5/2 state, for its unusual even denominator and possible application in topological quantum computation. Traditionally FQHE has been observed in high mobility GaAs heterostructure, but new materials such as graphene also open up a new area for FQHE. This review focuses on recent progress of FQHE at 5/2 state and FQHE in graphene.

**Keywords:** Fractional Quantum Hall Effect, Experimental Progress, 5/2 State, Graphene

## I. INTRODUCTION

### A. Quantum Hall Effect (QHE)

Hall effect was discovered in 1879, in which a Hall voltage perpendicular to the current is produced across a conductor under a magnetic field. Although Hall effect was discovered in a sheet of gold leaf by Edwin Hall, Hall effect does not require two-dimensional condition. In 1980, quantum Hall effect was observed in two-dimensional electron gas (2DEG) system [1,2]. QHE's voltage has a stepwise dependence on the magnetic field (B), as compared to a linear dependence as in Hall effect. Those step resistances are well defined plateaus, and quantized as $h/e^2$ divided by an integer number n, with extremely high resolution at the level of a few parts in $10^{10}$ [3,4]. Therefore QHE is also named as integer quantum Hall effect. In the meanwhile, longitudinal resistance becomes zero, if temperature is low enough, as shown in Fig. 1. Because quantum Hall plateaus have high resolution and only link two physics constants, QHE has been used as resistance calibration standard since 1990. In addition to its metrology application, QHE provides a new type of phase transition that cannot be described by Landau symmetry breaking theory.

In 2DEG, the eigenvalues of individual electron are quantized in Landau levels. The integer number n is the Landau level index if the spin energy splitting in a magnetic field is neglected. The ground state degeneracy is linear with B, and filling factor is defined as number of electrons over ground state degeneracy. The quantized Landau levels, filling factor dependence on B, together with the effect of disorder, cause the phenomenon of QHE. The delta-function-like Landau levels are expanded to localized states and extended states. When QHE happens, the Fermi level lies between extended states. The localized states created by impurities become the gaps for QHE and can be measured through the temperature dependence of longitudinal resistance. Due to the confinement potential of a realistic 2DEG sample, the gapped QHE state has chiral edge current at boundaries.

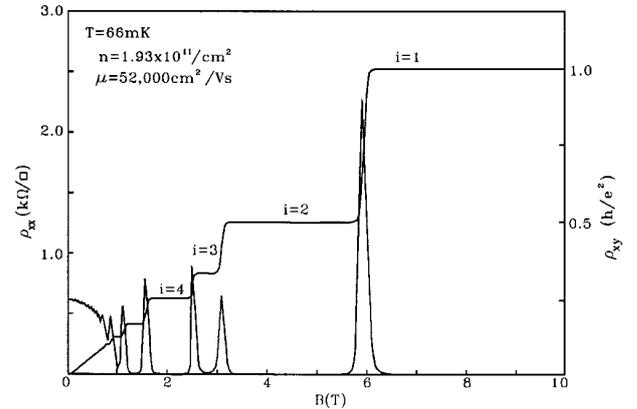

**Figure 1.** Hall resistance and longitudinal resistance in QHE. Adapted from [5].

The original QHE experiment was carried out on Si metal-oxide-semiconductor field effect transistors (MOSFET). A 2DEG is induced electrostatically by a metallic gate on top of $SiO_2$/Si. A modulation-doped GaAs-AlGaAs heterostructure can provide 2DEG and it is a better platform than Si MOSFET. For example, the best 2DEG in GaAs-AlGaAs heterostructure is over $3\times10^7$ $cm^2/(V\cdot s)$ in mobility [6,7]. The development of GaAs-AlGaAs heterostructure mobility can be found in Fig. 2. Electrons in GaAs-AlGaAs interface are confined in a potential well formed by the two semiconductors with band offset. All experiments mentioned in this review are carried on in 2DEG confined by GaAs-AlGaAs heterostructure unless otherwise specified.

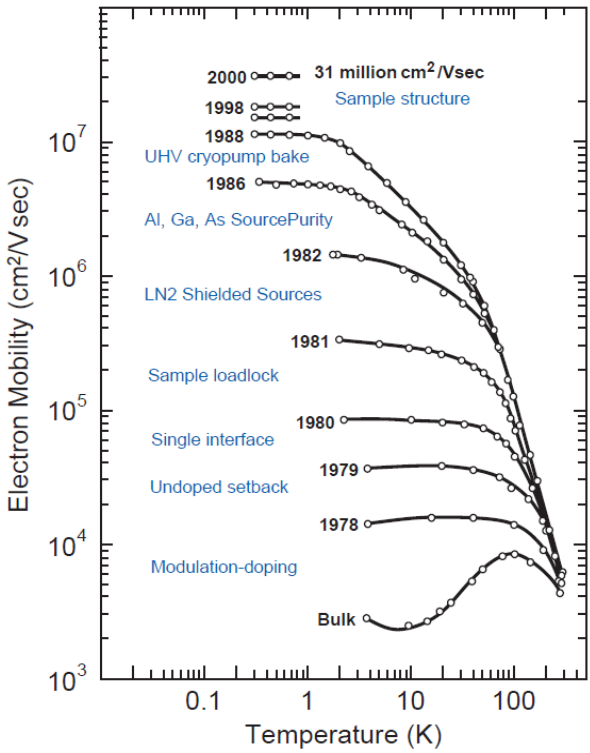

**Figure 2.** History of improvements in the mobility of 2DEG in GaAs-AlGaAs heterostructure. Adapted from [6].

**B. Fractional Quantum Hall Effect (FQHE)**

Two years after the first observation of QHE, FQHE was discovered by Tsui, Stormer and Gossard [8,9,5]. FQHE has almost the same characteristic as QHE, except that the quantized Hall resistance is $h/e^2$ divided by a fraction. The first fraction is 1/3 and around 100 fractional quantum Hall (FQH) states have been observed so far [10]. Most of them are odd denominator fractions. Higher mobility 2DEG and lower temperature are the keys to observed FQHE.

Laughlin proposed an elegant wave function to explain the first FQH state [11,12,5]. In this wave function, interaction between electrons is considered and Laughlin's wave function explains other 1/m (m is an odd integer) fractional quantum Hall states [11,12]. The similarity between QHE and FQHE requires energy gaps. Landau levels serve as the origin of gaps in QHE, and Laughlin's theory expects an energy gap in FQHE. It seems counterintuitive that numerous electrons form quasiparticle with charge less than a single electron, but Laughlin suggested excitations with fractional element charge as well. In particle physics, quarks are expected to carry 2e/3 or –e/3. In condensed matter physics, theoretical study in polymer proposed fractional charge at the domain boundary. However, FQHE is the first system actually observing fractional excitations. The 1/m states are predicted to have quasi-particles with e/m fractional charge. There has been plenty of evidence probing the fractional charge from transport [13], shot noise [14-16], interference [17-20], tunneling [21-24] and scanning single electron transistor [25,26] experiments.

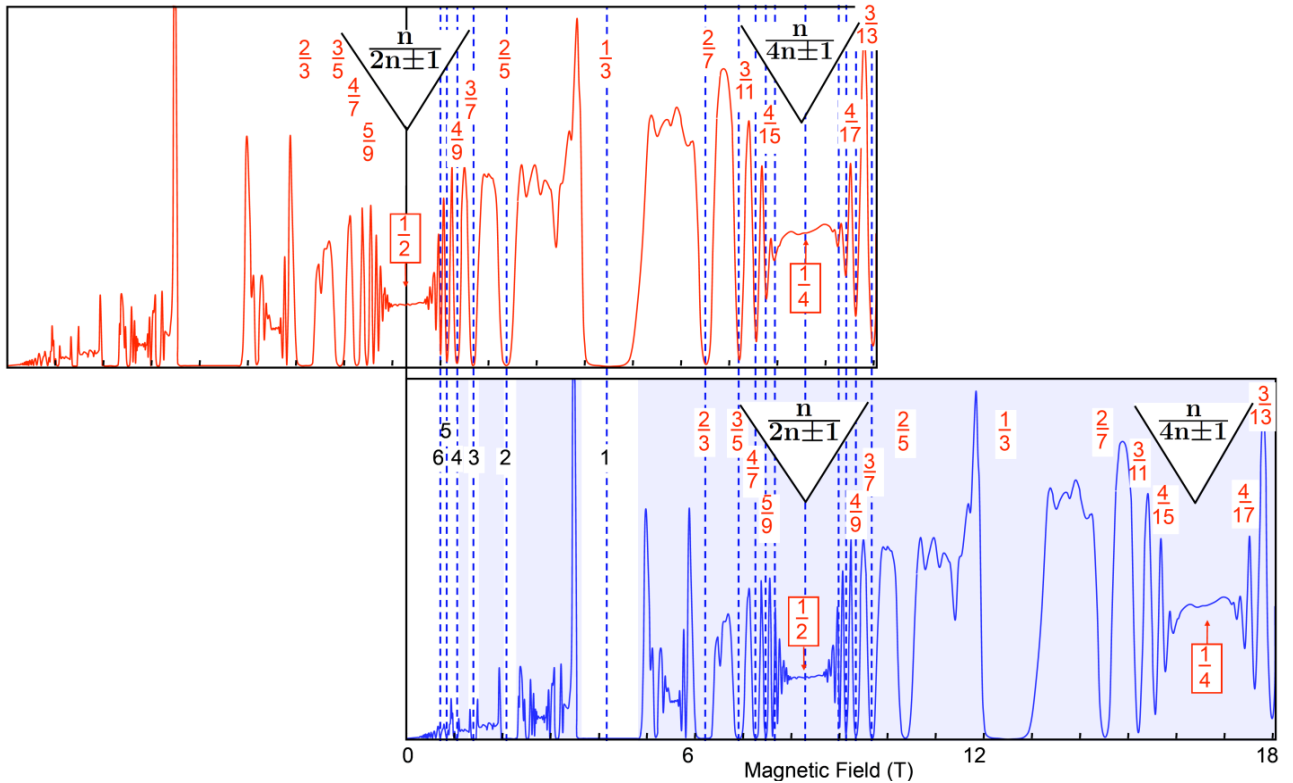

**Figure 3.** Demonstration of the similarity between FQHE and QHE (Source: H. L. Stormer). Adapted from [27].

Most of the fractional filling factors can be expressed as $\upsilon = n/(2pn \pm 1)$ and explained by the composite fermion theory [27]. The degeneracy per unit area is $B/\phi_0$ and $\phi_0 = h/e$, named as flux quantum. When the ground state is completely filled, namely at the ν=1 filling factor, the number of electron equates the number of flux. Composite fermion theory proposed by Jain suggests that each electron combines two flux, or an even number of flux, to form a composite fermion [28]. The new quasiparticle neglects interaction and moves in an effective magnetic field with $B^* = B - 2mn\phi_0$. The 2m represents the number of flux attached to an electron, and n is an analog to Landau level index. At the half filling factor, the effective magnetic field is equal to zero. Jain's picture translates a strongly correlated many body system to single quasiparticle physics. The composite fermions can be treated like electrons in QHE, as illustrated in Fig. 3 and Fig. 4. When the magnetic field deviates from 1/2 filling factor, composite fermions "feel" effective magnetic field in the same way as electrons "feel" real magnetic field in QHE. FQHE has edge current too, with the same reason as QHE does. The energy gap in FQHE can be measured similarly to QHE and energy gap of composite fermions can be grouped by number of flux attached to the electron [29].

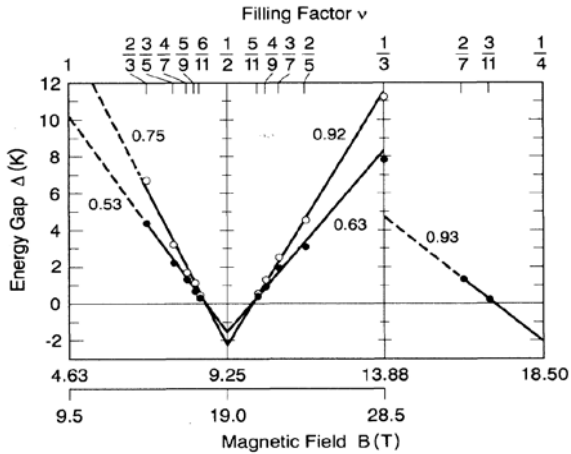

**Figure 4.** Energy gap at various filling factors in the vicinity of 1/2 and 1/4 filling factors. Adapted from [29].

QHE or FQHE states can be destroyed by current through 2DEG [30,31]. There is a critical current above which the Hall resistance deviates from quantized number and longitudinal resistance deviates from zero. It is reasonable to speculate that high current density causes dissipation and electron heating effect, and then destroys QHE. There are more theoretical candidates for the mechanism of breakdown, such as inter-Landau-level scattering, intrasubband process or electron-phonon interaction. A consensus of the exact explanation of QHE breakdown has not been reached, but all theories and measurements agree that critical current increases with less confinement condition. A recent experiment reports that FQHE has opposite dependence within a small confinement region, and calls for further theoretical input considering interactions [32]. Understanding the physics of breakdown is important for using the QHE Hall resistance plateau for metrology standard and helpful for some experiments with confinement described later in this review.

More reading about QHE or FQHE may refer to [33,34,28,35].

### C. Edge Physics and Statistics

All realistic 2DEG has a boundary, and edge states (or edge currents) originate from Landau energy gaps with the boundary potential [36]. Classical analog for a single edge current is electron move under Lorenz force, where only the electron near the edge can move forward without forming a close orbit [37]. Similar to its classical analog, edge currents move in different directions on two opposite boundaries. A real sample may have two boundaries separated by hundreds of microns to a few millimeters. Then two opposite edges are far from each other and backscattering is not allowed. Therefore, electron can move forward without dissipation, resulting in zero longitudinal resistance at the right filling factor and at sufficiently low temperature. The number of QHE edge states equates bulk Landau index n. The edge with the smallest index is closest to the boundary. The above argument applies to FQHE considering a composite fermion picture with similar energy gap. The experimental evidence of edge current can be found in non-local measurement [38], current distribution measurement [39] or direct imaging [40].

Although the bulk of 2DEG is gapped in QHE, the edge is gapless, where an electron underneath Fermi level can be excited above with infinitely small amount of energy. Wen pioneered in developing a chiral Luttinger liquid theory for edge excitation of FQHE [41-47]. The structure of edge can be complicated with counter-propagating modes and reconstruction in FQHE [48-50]. The history of chiral Luttinger liquid theory in FQHE can be found in [51]. Luttinger liquid theory predicts a tunneling exponent α for nonlinear I-V curve ($I \propto V^\alpha$), where I is tunneling current and V is tunneling voltage. This exponent represents an interaction parameter in Luttinger liquid which is related to Fermi energy and Coulomb energy. If edges are separated by vacuum then there is electron tunneling, and if edges are separated by FQH state then there is quasiparticle tunneling [52]. Electron tunneling can also happen between 2DEG and heavily doped GaAs contact on the side by cleaved-edge-overgrowth sample, structure shown in Fig. 5 [53,51].

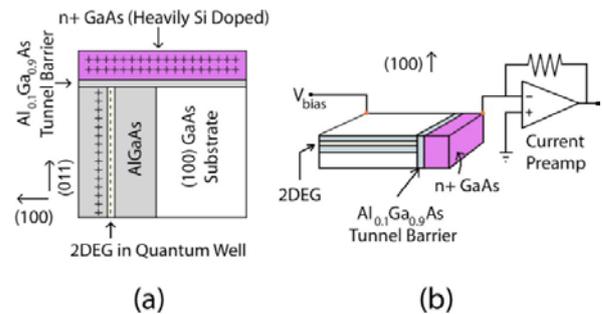

**Figure 5.** (a) Cleaved-edge-overgrowth device; (b) tunneling current measurement setup. Adapted from [51].

Exponents α for electron tunneling or quasiparticle tunneling are different for a given FQH state. Quasi-particle tunneling may happen between two counter propagating edge currents by bringing two edges appropriately close to each other. Particularly for further discussion in the next chapter, the zero-bias tunneling conductance is also in a power law dependence of temperature $\sigma \propto T^{2g-2}$. g is the strength of the Coulomb interaction and is also a topological quantum number independent of edge potential details. Different FQH states result in different g numbers so its measurement helps to determine wave function for theoretically solved states. For example, g is equal to 1/m for quasiparticle tunneling in a 1/m fractional Hall state (m is odd number). More about edge tunneling can be found in [53,54,51,55].

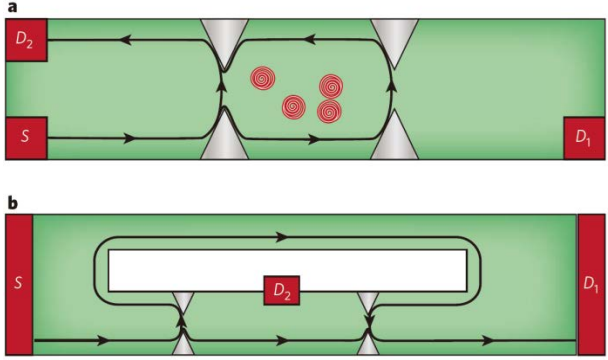

**Figure 6.** Interferometer experiment setups in QHE system, (a) for Fabry–Pérot interferometer and (b) for a Mach–Zehnder interferometer. Interferometer experiments mentioned in this review are based on Fabry–Pérot interferometer. Adapted from [56].

The g number is linked to the statistics of the particles, where statistics means the property of wave function under exchange. Wave functions in 3-dimensional space satisfy the symmetry property when two identical particles are exchanged. People are familiar with symmetric property for Bosons and anti-symmetric property for Fermions. The exchange of Bosons or Fermions generates a zero or π phase factor in the wave function, but in 2-dimensional space, this phase factor can be any value in principle. Such a quasiparticle with non-discrete phase value is called anyon [57-59]. The existence of fractional statistics can be derived by Laughlin's wave function in FQHE. Interference experiments (Fig. 6) based on edge physics are proposed for searching of fractional statistics [60]. Although one can argue that fractional statistics is a direct consequence of fractional charge, fractional statistics itself does not have as much experimental evidence as fractional charge in FQHE [61]. Some proposed waves functions are more complicated than what we describe above for fractional statistics, and the new statistics is named as non-Abelian statistics [57,62,59,56]. In non-Abelian statistics, interchange of particles not only changes the phase, but generates a completely different degenerate state. Non-Abelian statistics can be used for construction of quantum computer which is resistant to environmental decoherence, one of the major obstacles in quantum computation [57,62,63]. Filling factor 5/2 FQH state is a candidate for non-Abelian wave functions.

## II. 5/2: AN EVEN DENOMINATOR STATE

### A. The Mysterious 5/2

Among the known fractional states, 5/2 is one of the most mysterious states. 5/2 was first observed in 1987 by Willett et. al. (Fig. 7) [64]. Before the discovery of 5/2, all other FQH states have odd denominators. In the composite Fermion picture, the effective magnetic field is zero at filling 1/2. Even composite Fermion p-wave pairings may lead to an attractive interaction at 5/2 but not at 1/2 [65], 5/2 can be considered as a 1/2 at a higher Landau level. The existence of 5/2 contradicts the empirical observation from 1982 to 1987, and demands theoretical conceptual novelty.

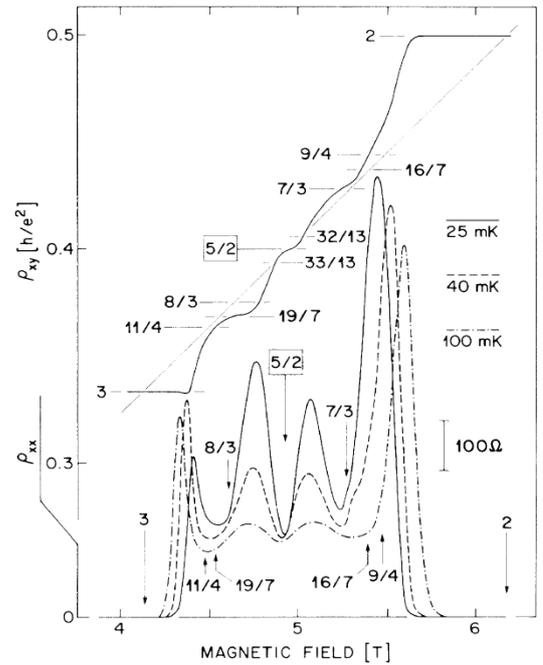

**Figure 7.** First experimental observation of 5/2 FQHE state. Adapted from [64].

Soon after the discovery of 5/2, the first theoretical attempt from Haldane and Rezayi [66] constructs quasiparticle by d-wave paring of composite fermions. It's suggested that Haldane-Rezayi's state has gapless excitations in the bulk [67]. Moore and Read proposed a theory for 5/2 in 1991, well known as the Moore-Read's state or Pfaffian state [68]. Moore-Read's state considers composite fermions as triplet paired, and the name Pfaffian came from the math used in this wave function. Numerical result from Morf in 1998 supports a spin polarized state like Pfaffian [69]. Anti-Pfaffian state is a particle-hole conjugate state to Pfaffian [70,71]. A way to distinguish between Pfaffian and anti-Pfaffian is through their different strength of the Coulomb interaction g. Both Pfaffian and anti-Pfaffian state are non-Abelian and so far they are believed to be the most likely wave function for 5/2 [69,72-84], with some supportive experimental evidence to be introduced in the following sections. Pfaffian state is also challenged by some numerical results, but some of them lead to a fractional charge of e/2, different from experimental value [85-89]. Halperin proposed 331 state for 5/2, similar to the 1/2 state in double layer system with two-component strongly paired Laughlin phases in each layer [90,91]. 331 state is Abelian and two recent tunneling measurements found g number closer to that of 331 state than others [23,24]. A numerical study compares spin unpolarized Halperin 331 state and the spin polarized Pfaffian state, and energetically favors Pfaffian state at ν=5/2 [84]. It should be noted that 331 state can be either spin-polarized or spin-unpolarized [92]. Wen also proposed U(1)×SU2(2) state (non-Abelian) and K=8 state (Abelian) in early 90s [93-95]. Feldman summarized some 5/2 wave functions including anti-331, anti K=8, anti-U(1)×SU2(2) and proposed a new 113 state [92,96]. Up till now, the existence of non-Abelian statistics at 5/2 and the theoretical explanation of 5/2 are still an open question experimentally.

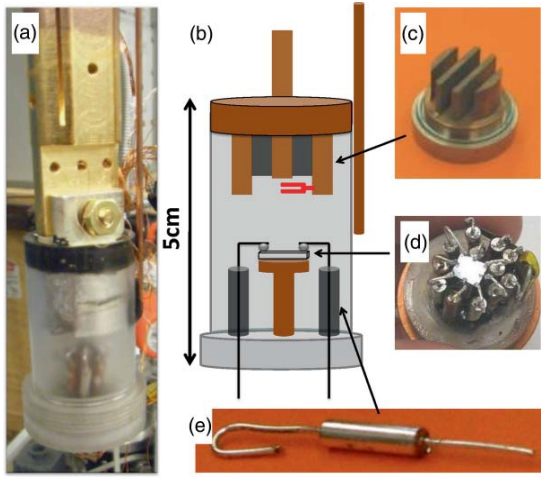

**Figure 8.** A design of $^3$He immersion cell. 2DEG in this kind of cell can reach 4 mK electron temperature. Adapted from [97]

5/2 is not the only even denominator FQHE. 7/2 and 19/8 filling factors have been observed with quantized Hall resistance and minimum in longitudinal resistance [98,99,10]. They attracted much less attention for their more special experimental requirements. For example, one 19/8 state's energy gap may be as low as 5 mK, requiring a low electron temperature environment that most of the lab cannot achieve [10]. The temperature mentioned here is electron temperature rather than the environmental lattice temperature provided by a cryostat. A dilution fridge can go down as low as 2 mK and a nuclear demagnetization refrigerator can go down to the order of 10 μK lattice temperature, but only a few labs achieve electron temperature lower than 10 mK [100,97] A design of $^3$He immersion cell achieving 4 mK is shown in Fig. 8. For comparison, although 5/2 state is considered extremely difficult to observe, its energy gap can be as high as 0.5 K, 100 times larger than 19/8 state. 9/2 and 11/2 filling factors in 2DEG are studied for their anisotropic transport property [101-104]. Anisotropic transport in 2D hole systems is slightly different [105,106]. The half filling states show novel behaviors from the lowest Landau level to higher Landau level. Higher Landau levels suppress the short range Coulomb repulsions between electrons and cause the anisotropic stripe phase or bubble phase. The only two known FQHE at ν>4 are 21/5 and 24/5, and they only exist between 80 and 120 mK [107].

### B. Energy Gap of 5/2

5/2 FQH state is difficult to obverse due to the requirement of high mobility and low temperature. Energy gap indicates how robust an FQH state is. 5/2 state's energy gap is much smaller than the usual FQH state such as 1/3. The observed 5/2 energy gap is from less than 0.6 K to undetectable. In comparison, 1/3's energy gap has been measured up to ~10 K. [29] and numerical study predicts that 1/3 state has 11 times larger energy gap than 5/2 in the same sample [69]. Traditionally 5/2 is believed to only occur in the extremely high mobility samples, usually in the order of $10^7$ cm$^2$/(V·s). A recent study found 5/2 in a sample with mobility as low as $4.8\times10^6$ cm$^2$/(V·s) [108]. More experimental results show that energy gap has complicated dependence on mobility, density and other properties of the heterostructure.

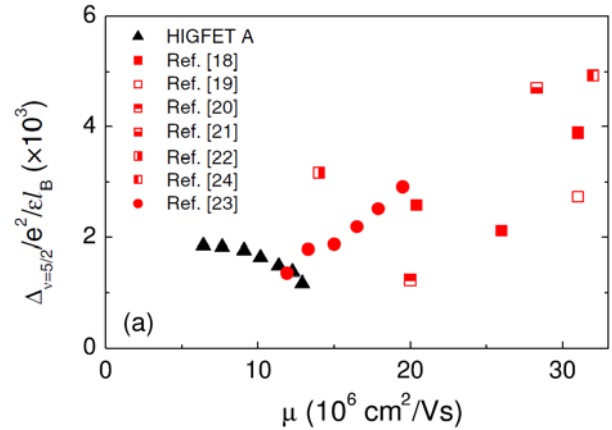

**Figure 9.** A summary of 5/2 reduced energy gap as a function of mobility. Black triangle data are from an undoped heterojunction insulated gate field-effect transistors. Adapted from [109].

In general, samples from the same batch have larger energy gap with higher mobility (Fig. 9) [109,108]. Mobility dependence on undoped heterojunction insulated gate field-effect transistors shows that energy gap slightly increases with lower mobility. One should notice that the mobility itself is also a function of 2DEG electron density, either monotonically increasing or non-monotonic [110,109]. The density dependence of the energy gap seems consistent in that larger density leads to larger energy gap, especially in the same study [110,109,111]. A 5/2 study in wide quantum well also found that energy gap increases with density but it suddenly drops once the density exceeds a number associated electric subband switch [112]. The role of disorder and Landau level mixing has been discussed for explaining the measured 5/2 energy gap [113,114,7,110,109,115,108,116]. Disorder is not included in the 5/2 ground state numerical simulations to our knowledge. Scanning gate microscopy or scanning probes can serve as detectors for individual hard-scattering site and disorder potential [117-119]. A 2-impurity model (background and remote) was developed recently to address the relation between mobility, electron density, disorder and sample quality [120,121]. In addition to mobility and density, conditions for LED illumination, Aluminum fractions and silicon doping levels in heterostructure also affect the 5/2 FQH state [108,116].

### C. Spin Polarized or Unpolarized?

Numerical results suggest that 5/2 FQH state is spin polarized and support Pfaffian state or anti-Pfaffian state [69,72,77,80,82]. Therefore, efforts have been made to measure the spin polarization at 5/2, mostly through tilted field technique. In tilted field measurement, the Hall properties refer to perpendicular magnetic field while the Zeeman energy refers to the total magnetic field, so a spin polarized state will not be affected by the tilted field while a spin unpolarized state may respond to the tilted field. The first 5/2 tilted field experiment was done by Eisenstein in 1988 [122] and the study concluded that 5/2 is not spin polarized. More titled experiments were carried out during the last decade after more theoretical support for the Pfaffian and anti-Pfaffian, but the results were too complicated to be explained only in language of spin polarization [103,123-129]. The above titled experimental results imply that 5/2 state under the influence of in plane field needs to take more into consideration than just the Zeeman energy. Furthermore, a numerical study argues that 5/2 from Moore-Read state can lead to depolarized ground state in realistic experimental situations because of Skyrmions [130].

NMR or optical measurement can also provide insight on spin polarization. Experimental results are more in favor of spin polarized state than unpolarized state [131-135]. A recent study points out that 5/2 may not be spin polarized in the low-density limit through the density dependence of energy gap [136]. In addition, another analysis through measurement of energy gap shows that 5/2 is more consistent with spin unpolarized than spin polarized [137]. In short, although spin polarization provides valuable information for determining wave functions, it is difficult to make one conclusion for all the samples measured. Readers may refer to [138-140] for more information about spin polarization and tilted field results.

### D. e/4 Fractional Charge and Interaction Parameter g

Fractional charge e/4 predicted by Pfaffian, anti-Pfaffian, 331 or other wave functions has been observed by different experimental methods [89,22,18,141,19,26,23,24]. Shot noise happens when discontinuous current passes through a barrier, with a magnitude related to the current, the quantized charge unit, and the barrier itself. The barrier is basically a narrow constriction created through a quantum point contact (QPC), where edge currents from both sides of a sample are brought close to each other. The formation of QPC can be through top gates capacitively coupled to the 2DEG or through etching away selective area of a 2DEG. Shot noise experiment proves the existence of e/4 quasiparticle (Fig. 10) at 5/2 and fractional charge at 5/2 varying with temperature [89,141,142]. The observation of e/4 is consistent with theoretical predictions but it is not enough to conclude the existence of non-Abelian statistics. Neutral mode at 5/2 has been measured in shot noise experiment and explained in support of non-Abelian statistics [143]. A recent theoretical study suggests a new Abelian 5/2 state that allows neutral mode [96]. Neutral mode has been observed in other FQH states, such as 2/3, 3/5, 5/3, 8/3 recently [144].

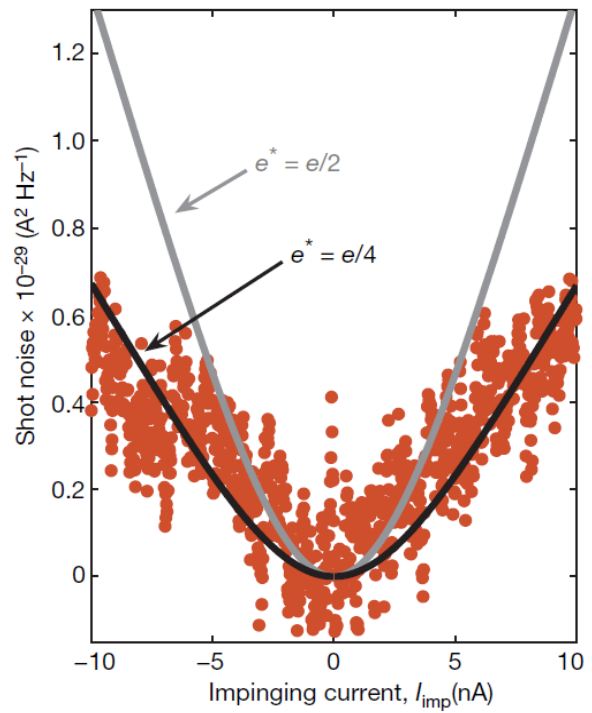

**Figure 10.** e/4 fractional charge detected from shot noise measurement. Adapted from [89].

QPC can also be used in tunneling experiments where edge currents are close enough to have backscattering. Although it was initially developed in Abelian FQHE, tunneling theory applies to non-Abelian states and tunneling itself as a measurement tool has been studied theoretically and experimentally [52,47,54,145-157,92]. Weak tunneling experiment observes fractional charge close to e/4 at 5/2 FQH state, also provides the information of interaction parameter $g$ [22,158,23,24]. Different proposed wave functions predict several numbers of $g$ and it can be measured through weak tunneling theory [93,43,45,46,52,151]. The first weak tunneling experiment at 5/2 found $g$=0.35 in its best fit compared with the weak tunneling formula (Fig. 11) [22]. The improved measurement from the same group observed fractional charge close to 0.25e and suggested that g number is closer to 3/8, predicted by Abelian 331 wave function [52,23]. Minima on both sides of the zero-bias tunneling peak indicates that g is strictly less than 1/2 [159,23]. Theoretical efforts point out the possibility of geometry influence in this kind of QPC tunneling measurement and g dependence on edge reconstruction [160,92]. A recent tunneling measurement with a different grown sample and different electron density from ETH also supports 331 Abelian state and provides tunneling information at 7/3 and 8/3 for further theoretical investigation [24,161].

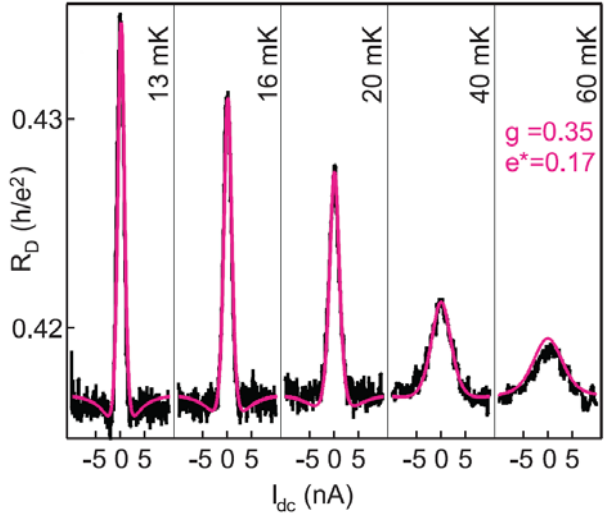

**Figure 11.** Determination of fractional charge and interaction parameter g through weak tunneling experiment. Adapted from [22].

Interferometer has been proposed to detect fractional charge and fractional statistics by Chamon [60]. Furthermore, interference measurement can be used to prove the existence of non-Abelian statistics directly and there are considerable theoretical efforts in interference study [162-182]. It is predicted that the interference would be turned on and off depending on if the non-Abelian quasiparticles within the interferometer are even or odd number. The simplest interferometer can be constructed with two QPCs (Fig. 6a). Willett's interference experiment observed fractional charge of e/4 and e/2 and are considered in support of the non-Abelian statistics [18,19,183]. In addition to the fractional charge, an alternating pattern of e/4 and e/2 period oscillation is present, similar to the even-odd effect predicted for non-Abelian quasiparticles. The e/2 period may reflect a more complicated situation of e/4 quasiparticle around an interferometer and the interference results can be a signature of non-Abelian [184-186]. The physics of interference in 2DEG is more than just the resistance oscillations, for another study proves that the oscillations come from Aharonov-Bohm mechanism in a larger interferometer and from Coulomb blockade mechanism in a smaller interferometer [187]. There is one more 5/2 interferometry experiment supporting non-Abelian statistics through the detection of phase slips [188].

In short, FQH 5/2 state needs extra efforts to determine its ground state wave function. So far experimental results are in more accordance with the non-Abelian nature of 5/2 than the Abelian nature. Confinement potential can affect the wave function of 5/2 FQHE from one numerical study [184]. Therefore, given the results from energy gap study, spin polarization probes, and experiments carried out with QPC, device geometry and heterostructure characteristics may be important.

### III. FQHE IN GRAPHENE

#### A. Graphene

Electrons and their interaction fantastically result in FQHE, regardless how the 2DEG is formed. Besides Si-MOSFET and GaAs-AlGaAs heterostructure, other 2DEG systems are also used in QHE study. For example, 2DEG based on MgZnO/ZnO heterostructure reached the mobility of $7 \times 10^5$ cm$^2$/(V·s) only 4 years after its first demonstration as a host for 2DEG [189,190] and many FQH states have been observed in ZnO system [191,192]. FQHE has been also observed in Si/SiGe heterostructure [193,194] with mobility in the order of $10^6$ cm$^2$/(V·s) nowadays. Besides the familiar heterostructure, two-dimensional materials can also host FQHE. In fact, there are a large number of two-dimensional materials, among which graphene is frequently studied [195-197].

Graphene is special for its peculiar band structure and corresponding two-dimensional massless Dirac-like excitations. Landau levels in regular 2DEG are characterized as separation of $\hbar\omega_C$, where $\omega_C = eB/m^*$ and m* is effective mass. Obviously, graphene cannot adapt regular Landau levels from standard 2DEG. Eigenvalues in monolayer graphene are expressed as $E_n = \text{sgn}(n)v_F\sqrt{2e\hbar B|n|}$. There is a zero-energy Landau level in graphene, qualitatively different from standard 2DEG. Spin splitting g number is 2 in graphene and -0.44 in GaAs semiconductor.

#### B. QHE in Graphene

In graphene, Hall conductance is $\sigma_H = \pm 4(n+1/2)e^2/h$. The factor 4 results from the spin degeneracy and valley degeneracy. The additional 1/2 can natively attribute to the existence of the zero energy state. More reading about this additional 1/2 is available in [196] from the view of Berry phase or from the view of analogy to relativistic Dirac equation.

Hall plateaus in graphene develop to the series of filling factors at $\upsilon = \pm 2, \pm 6, \pm 10 \cdots$ (shown in Fig. 12). From 2005, the predicted graphene quantum Hall series have been experimentally observed [198,199]. $\upsilon = 0, \pm 1, \pm 3$ have been experimentally observed from symmetry broken of the spin or valley [200-202]. Due to its much larger energy gap than that in standard 2DEG, QHE in graphene has even been found at room temperature [203], which makes QHE based applications more practical in the future. With the developing device fabrication process, more sophisticated structures have been realized which enable further investigation of edge current in graphene [204-206].

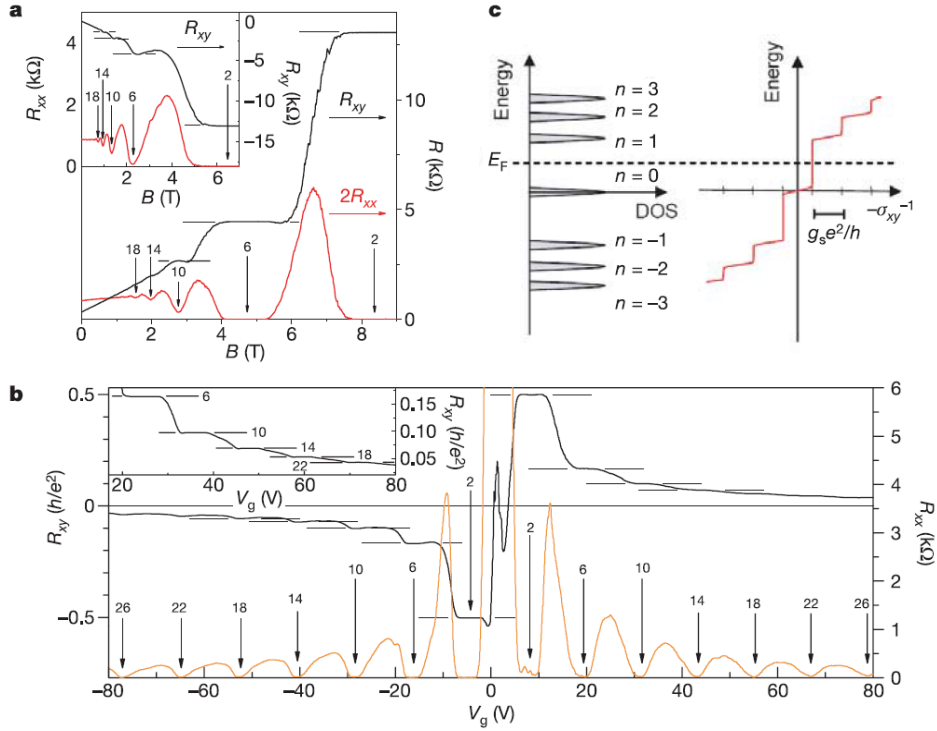

**Figure 12.** QHE in graphene. (a) Hall resistance and longitudinal resistance measured at 30 mK and $V_g$=15 V. (b) Hall resistance and longitudinal resistance at 1.6 K and 9 T. (c) A schematic diagram of the Landau level density of states and corresponding quantum Hall conductance as a function of energy. Adapted from [199].

### C. FQHE in Graphene

FQHE in graphene was observed 4 years after the discovery of QHE in graphene, longer than the time difference of QHE in 1980 and FQHE in 1982 [201,207]. The early 1/3 FQHE was found in suspended graphene devices at a temperature of 1.2 K and a magnetic field of 12 T, where the mobility reached $2.6 \times 10^5$ cm$^2$/(V·s) [201]. Carrier mobility in suspended graphene can be in the order of $10^6$ cm$^2$/(V·s), comparable to some good quality 2DEG in GaAs-AlGaAs heterostructure [208]. Substrate affects the properties of graphene so suspended graphene has its own advantages. However, suspended graphene is unstable, inconvenient for further complicated fabrications and easy to bring in disorder from strain [209,210]. High quality graphene on a hexagonal boron nitride (h-BN) substrate was found and Hall bar structure was realized [211]. In Fig. 13, at least 8 FQHE states were found in graphene on h-BN substrate, at 0.3 K and 35 T, where the mobility was $3 \times 10^4$ cm$^2$/(V·s) [202]. The FQHE states in monolayer graphene in reference [207,201,202,212-215] can be summarized in the following tables:

| 1/3 | 1/5 | 2/7 | 2/9 | 3/11 | | | | | | |
|---|---|---|---|---|---|---|---|---|---|---|
| 2/3 | 2/5 | 3/7 | 4/9 | 5/11 | | | | | | |
| 4/3 | 3/5 | 4/7 | 5/9 | | | | | | | |
| 7/3 | 6/5 | 5/7 | 14/9 | | | | | | | |
| 8/3 | 8/5 | 10/7 | | | | | | | | |
| 10/3 | | | | | | | | | | |
| 11/3 | | | | | | | | | | |
| 13/3 | | | | | | | | | | |
| | | | | | | | | | | |
| 1/3 | 1/5 | 1/7 | 1/9 | 2/11 | 2/13 | 2/15 | 2/17 | 3/19 | 5/21 | 6/23 | 6/25 |
| 8/3 | 24/5 | 19/7 | 25/9 | 17/11 | 20/13 | 23/15 | 9/17 | 10/19 | 10/21 | | |

**Table 1.** FQHE in monolayer graphene (black color). Red color fractions are odd denominator FQHE states in GaAs-AlGaAs heterostructure [10]. The first red color number in each column is the lowest filling factor with a particular odd denominator. The second red color number, if it exists, is the highest filling factor with a particular odd denominator.

| Reference | Observed fractions | T (K) | B (T) | Mobility ($\times 10^5$ cm$^2$/(V·s)) | Notes |
|---|---|---|---|---|---|
| [201] | 1/3 | 1.2 | 12 | 2.6 | Suspended graphene |
| [207] | 1/3 | 2 | 14 | 1 | Suspended graphene |
| [212] | 1/3 | 1.7 | 14 | 1.5 | Suspended graphene with Hall Bar |
| [202] | 1/3, 2/3, 4/3, 7/3, 8/3, 10/3, 11/3, 13/3 | 0.3 | 35 | 0.3 | Hexagonal boron-nitride substrate with Hall Bar |
| [214] | 1/3, 2/3, 2/5, 3/7, 4/9 | 1.4 | 15 | Transconductance measurement | Representative data stem from suspended graphene |
| [213] | 1/3, 2/3, 4/3, 2/5, 3/5, 8/5, 3/7, 4/7, 10/7, 4/9, 14/9 | 0.45 | 12 | Local electronic compressibility measurement | Suspended graphene |
| [215] | 1/3, 2/3, 1/5, 2/5, 3/5, 6/5, 2/7, 3/7, 4/7, 5/7, 2/9, 4/9, 5/9, 3/11, 5/11 | N/A | 12 | Local electronic compressibility measurement | Suspended graphene |

**Table 2.** Summary of FQHE experimental studies in monolayer graphene.

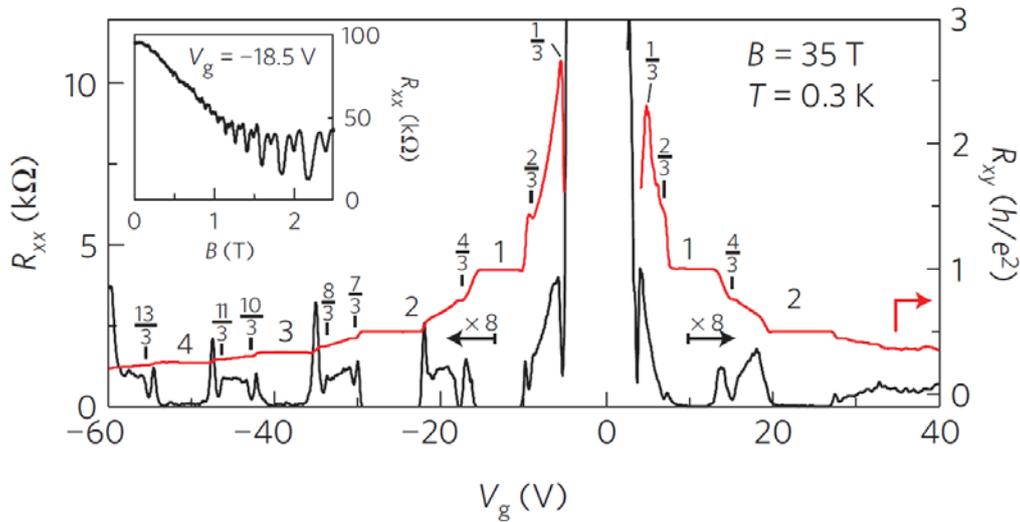

**Figure 13.** Hall resistance and longitudinal resistance at 35 T and 0.3 K. Inset shows SdH oscillations at $V_g$=-18.5 V. Adapted from [202].

Based on the summary of existing FQHE in graphene and standard 2DEG, the main difference does not origin from the absence of larger odd denominator states in graphene. With lower temperature and higher quality samples, one would expect more FQHE states' appearance in monolayer graphene. The existence of 13/3 and the lack of 5/3 indicate some differences. Electron-electron interaction is suppressed at higher Landau level so FQHE ground state may be replaced by stripe phase [102] or bubble phase [104]. In fact, there is no FQHE ground state above second Landau level experimentally found in GaAs-AlGaAs heterostructure. The 24/5 state listed in Table 1 is not the ground state, and it only shows up at temperature from 80 to 120 mK [107]. Graphene enables the FQHE at higher Landau levels, leading to the study of different effective interactions from different structures of Landau level wave functions. For example, in the second or higher Landau level, lots of interesting phenomena has been observed in GaAs standard 2DEG, such as even denominator states and reentrant states. The lack of 5/3 state relates to the degeneracy difference between that in graphene and GaAs. The symmetry in graphene may be broken by different experimental approaches, such as tilted field or strain. Furthermore, edge study in standard 2DEG is difficult for the 2DEG laying a few hundred nm below the surface, so the experimental definition of edge is usually through the etching of Mesa or capacitively coupled gating, which either cause disorder or can only define very smooth potential gradience. In comparison, edges in graphene can have atomically sharp confinement, leading to more studies of edge current tunneling and interferometer. In short, FQHE in graphene provides an interesting platform for many-body physics.

## IV. CONCLUSION

Many theoretical and experimental efforts are still ongoing after 33 years of FQHE research. The recent progresses at filling factor 5/2 state and FQHE study in graphene are briefly reviewed. 5/2 state has potential applications for topological quantum computation. The existing experimental inconsistencies imply that 5/2 could have different wave functions depending on their host environment. Even without non-Abelian statistics, 5/2 state itself is also important for being an even denominator state. This even denominator FQH state provides an exceptional platform for us to study the complicated many body physics from simple electrons. The progress of FQHE study in monolayer graphene is briefly introduced. FQHE states observed in traditional transport measurements are still limited to denominator 3. Temperature for graphene FQHE study is much higher than typical FQHE study in GaAs heterostructure. With improving monolayer graphene sample quality and lower measurement temperatures, more FQHE states are expected to be observed, likely including even denominator states and perhaps non-Abelian statistics. Graphene can be investigated for high Landau level FQHE and edge physics.

We write this brief review as the token of our honest regard for the many brilliant and serious minds working in this exciting field. This paper is essentially a review for the latest experimental progress at 5/2 and that of FQHE in graphene. Due to the framework of this article, some recent inspiring FQHE experiments are not introduced here, including coherence length measurement [216], heat transport [217], local electronic state measurement [218], neutral mode detection by local thermometry [219], nuclear spin polarization [220], piezoresistance [221-223], properties of FQHE in wide quantum well [224,112,225,226], reentrant [227,104], thermopower experiment [228,229], even denominator FQHE in bilayer graphene [230], FQHE in monolayer-bilayer graphene planar junctions [231] Hofstadter's butterfly in graphene [232], tunneling spectroscopy of graphene FQHE [233]. We apologize if any important related work is neglected because of our unawareness.

**FUNDING**

This work is financially supported by NBRPC (Grant No. 2012CB821402 and 2012CB921301) and NSFC (Grant No. 91221302, 11274020 and 11322435).

# REFERENCES

1. Klitzing, Kv, Dorda, G, Pepper, M. New method for high-accuracy determination of the fine-structure constant based on quantized Hall resistance. *Phys Rev Lett* 1980; **45**: 494.
2. Klitzing, Kv. The quantized Hall effect. *Rev Mod Phys* 1986; **58**: 519.
3. Jeckelmann, B, Jeanneret, B, Inglis, D. High-precision measurements of the quantized Hall resistance: Experimental conditions for universality. *Phys Rev B* 1997; **55**: 13124.
4. Jeckelmann, B, Jeanneret, B. The quantum Hall effect as an electrical resistance standard. *Séminaire Poincaré* 2004; **2**: 39.
5. Tsui, DC. Nobel Lecture: Interplay of disorder and interaction in two-dimensional electron gas in intense magnetic fields. *Rev Mod Phys* 1999; **71**: 891.
6. Pfeiffer, L, West, KW. The role of MBE in recent quantum Hall effect physics discoveries. *Physica E* 2003; **20**: 57.
7. Umansky, V, Heiblum, M, Levinson, Y, *et al.* MBE growth of ultra-low disorder 2DEG with mobility exceeding $35\times10^6 cm^2/(V\cdot s)$. *J Cryst Growth* 2009; **311**: 1658.
8. Tsui, DC, Stormer, HL, Gossard, AC. Two-dimensional magnetotransport in the extreme quantum limit. *Phys Rev Lett* 1982; **48**: 1559.
9. Stormer, HL. Nobel Lecture: The fractional quantum Hall effect. *Rev Mod Phys* 1999; **71**: 875.
10. Pan, W, Xia, JS, Stormer, HL, *et al.* Experimental studies of the fractional quantum Hall effect in the first excited Landau level. *Phys Rev B* 2008; **77**: 075307.
11. Laughlin, RB. Anomalous quantum Hall effect: an incompressible quantum fluid with fractionally charged excitations. *Phys Rev Lett* 1983; **50**: 1395.
12. Laughlin, RB. Nobel Lecture: Fractional quantization. *Rev Mod Phys* 1999; **71**: 863.
13. Clark, RG, Mallett, JR, Haynes, SR, *et al.* Experimental determination of fractional charge e/q for quasiparticle excitations in the fractional quantum Hall effect. *Phys Rev Lett* 1988; **60**: 1747.
14. de-Picciotto, R, Reznikov, M, Heiblum, M, *et al.* Direct observation of a fractional charge. *Nature* 1997; **389**: 162.
15. Saminadayar, L, Glattli, DC, Jin, Y, *et al.* Observation of the e/3 fractionally charged laughlin quasiparticle. *Phys Rev Lett* 1997; **79**: 2526.
16. Reznikov, M, Picciotto, Rd, Griffiths, TG, *et al.* Observation of quasiparticles with one-fifth of an electron's charge. *Nature* 1999; **399**: 238.
17. Camino, FE, Zhou, W, Goldman, VJ. e/3 Laughlin quasiparticle primary-filling ν=1/3 interferometer. *Phys Rev Lett* 2007; **98**: 076805.
18. Willett, RL, Pfeiffer, LN, West, KW. Measurement of filling factor 5/2 quasiparticle interference with observation of charge e/4 and e/2 period oscillations. *Proc Natl Acad Sci USA* 2009; **106**: 8853.
19. Willett, RL, Pfeiffer, LN, West, KW. Alternation and interchange of e/4 and e/2 period interference oscillations consistent with filling factor 5/2 non-Abelian quasiparticles. *Phys Rev B* 2010; **82**: 205301.
20. McClure, DT, Chang, W, Marcus, CM, *et al.* Fabry-Perot Interferometry with Fractional Charges. *Phys Rev Lett* 2012; **108**: 256804.
21. Goldman, VJ, Su, B. Resonant tunneling in the quantum Hall regime: measurement of fractional charge. *Science* 1995; **267**: 1010.
22. Radu, IP, Miller, JB, Marcus, CM, *et al.* Quasi-particle properties from tunneling in the ν=5/2 fractional quantum Hall state. *Science* 2008; **320**: 899.
23. Lin, X, Dillard, C, Kastner, MA, *et al.* Measurements of quasiparticle tunneling in the ν=5/2 fractional quantum Hall state. *Phys Rev B* 2012; **85**: 165321.
24. Baer, S, Rössler, C, Ihn, T, *et al.* Experimental probe of topological orders and edge excitations in the second Landau level. *arXiv: 1405.0428* 2014.
25. Martin, J, Ilani, S, Verdene, B, *et al.* Localization of fractionally charged quasi-particles. *Science* 2004; **305**: 980.
26. Venkatachalam, V, Yacoby, A, Pfeiffer, L, *et al.* Local charge of the ν=5/2 fractional quantum Hall state. *Nature* 2011; **469**: 185.
27. Jain, JK. A note contrasting two microscopic theories of the fractional quantum Hall effect. *arXiv: 1403.5415* 2014.
28. Jain, JK. *Composite fermions*. Cambridge, UK: Cambridge University Press; 2007.
29. Du, RR, Stormer, HL, Tsui, DC, *et al.* Experimental evidence for new particles in the fractional quantum Hall effect. *Phys Rev Lett* 1993; **70**: 2944.
30. Watts, JP, Usher, A, Matthews, AJ, *et al.* Current breakdown of the fractional quantum Hall effect through contactless detection of induced currents. *Phys Rev Lett* 1998; **81**: 4220.
31. Nachtwei, G. Breakdown of the quantum Hall effect. *Physica E* 1999; **4**: 79.
32. Dillard, C, Lin, X, Kastner, MA, *et al.* Breakdown of the integer and fractional quantum Hall states in a quantum point contact. *Physica E* 2013; **47**: 290.
33. Prange, RE, Girvin, SM. *The quantum Hall effect*: Springer New York; 1990.
34. Das Sarma, S, Pinczuk, A. *Perspectives in quantum Hall effects*. New York: John Wiley & Sons, Inc; 1997.
35. Ezawa, ZF. *Quantum Hall effects: recent theoretical and experimental developments*: World Scientific Publishing Company Incorporated; 2013.
36. Halperin, BI. Quantized Hall conductance, current-carrying edge states, and the existence of extended states in a two-dimensional disordered potential. *Phys Rev B* 1982; **25**: 2185.

37. Beenakker, CWJ, van Houten, H. Quantum transport in semiconductor nanostructures. *Solid State Phys* 1991; **44**: 1.
38. McEuen, PL, Szafer, A, Richter, CA, *et al.* New resistivity for high-mobility quantum Hall conductors. *Phys Rev Lett* 1990; **64**: 2062.
39. Fontein, PF, Kleinen, JA, Hendriks, P, *et al.* Spatial potential distribution in GaAs/Al$_x$Ga$_{1-x}$As heterostructures under quantum Hall conditions studied with the linear electro-optic effect. *Phys Rev B* 1991; **43**: 12090.
40. Lai, K, Kundhikanjana, W, Kelly, MA, *et al.* Imaging of Coulomb-riven quantum Hall edge states. *Phys Rev Lett* 2011; **107**: 176809.
41. Wen, XG. Chiral Luttinger liquid and the edge excitations in the fractional quantum Hall states. *Phys Rev B* 1990; **41**: 12838.
42. Lee, DH, Wen, XG. Edge excitations in the fractional-quantum-Hall liquids. *Phys Rev Lett* 1991; **66**: 1765.
43. Wen, XG. Edge transport properties of the fractional quantum Hall states and weak-impurity scattering of a one-dimensional charge-density wave. *Phys Rev B* 1991; **44**: 5708.
44. Wen, XG. Edge excitations in the fractional quantum hall states at general filling fractions. *Mod Phys Lett B* 1991; **05**: 39.
45. Wen, XG. Theory of the edge states in fractional quantum hall effects. *Int J Mod Phys B* 1992; **06**: 1711.
46. Chamon, CdC, Wen, XG. Resonant tunneling in the fractional quantum Hall regime. *Phys Rev Lett* 1993; **70**: 2605.
47. Wen, XG. Topological orders and edge excitations in fractional quantum Hall states. *Adv Phys* 1995; **44**: 405.
48. MacDonald, AH. Edge states in the fractional-quantum-Hall-effect regime. *Phys Rev Lett* 1990; **64**: 220.
49. Johnson, MD, MacDonald, AH. Composite edges in the ν=2/3 fractional quantum Hall effect. *Phys Rev Lett* 1991; **67**: 2060.
50. Chamon, CdC, Wen, XG. Sharp and smooth boundaries of quantum Hall liquids. *Phys Rev B* 1994; **49**: 8227.
51. Chang, AM. Chiral Luttinger liquids at the fractional quantum Hall edge. *Rev Mod Phys* 2003; **75**: 1449.
52. Wen, XG. Topological order and edge structure of ν=1/2 quantum Hall state. *Phys Rev Lett* 1993; **70**: 355.
53. Chang, AM, Pfeiffer, LN, West, KW. Observation of chiral Luttinger behavior in electron tunneling into fractional quantum Hall edges. *Phys Rev Lett* 1996; **77**: 2538.
54. Milliken, FP, Umbach, CP, Webb, RA. Indications of a Luttinger liquid in the fractional quantum Hall regime. *Solid State Commun* 1996; **97**: 309.
55. Wen, XG. *Quantum field theory of many-body systems: from the origin of sound to an origin of light and electrons*. Oxford: Oxford University Press; 2007.
56. Stern, A. Non-Abelian states of matter. *Nature* 2010; **464**: 187.
57. Kitaev, AY. Fault-tolerant quantum computation by anyons. *Ann Phys* 2003; **303**: 2.
58. Weeks, C, Rosenberg, G, Seradjeh, B, *et al.* Anyons in a weakly interacting system. *Nat Phys* 2007; **3**: 796.
59. Stern, A. Anyons and the quantum Hall effect—A pedagogical review. *Ann Phys* 2008; **323**: 204.
60. Chamon, CdC, Freed, DE, Kivelson, SA, *et al.* Two point-contact interferometer for quantum Hall systems. *Phys Rev B* 1997; **55**: 2331.
61. Camino, FE, Zhou, W, Goldman, VJ. Realization of a Laughlin quasiparticle interferometer: Observation of fractional statistics. *Phys Rev B* 2005; **72**: 075342.
62. Nayak, C, Simon, SH, Stern, A, *et al.* Non-Abelian anyons and topological quantum computation. *Rev Mod Phys* 2008; **80**: 1083.
63. Bonderson, P, Feiguin, AE, Nayak, C. Numerical calculation of the neutral fermion gap at the ν=5/2 fractional quantum Hall state. *Phys Rev Lett* 2011; **106**: 186802.
64. Willett, R, Eisenstein, JP, Störmer, HL, *et al.* Observation of an even-denominator quantum number in the fractional quantum Hall effect. *Phys Rev Lett* 1987; **59**: 1776.
65. Scarola, VW, Park, K, Jain, JK. Cooper instability of composite fermions. *Nature* 2000; **406**: 863.
66. Haldane, FDM, Rezayi, EH. Spin-singlet wave function for the half-integral quantum Hall effect. *Phys Rev Lett* 1988; **60**: 956.
67. Read, N, Green, D. Paired states of fermions in two dimensions with breaking of parity and time-reversal symmetries and the fractional quantum Hall effect. *Phys Rev B* 2000; **61**: 10267.
68. Moore, G, Read, N. Nonabelions in the fractional quantum Hall effect. *Nucl Phys B* 1991; **360**: 362.
69. Morf, RH. Transition from quantum Hall to compressible states in the second landau level: new light on the ν=5/2 enigma. *Phys Rev Lett* 1998; **80**: 1505.
70. Lee, SS, Ryu, S, Nayak, C, *et al.* Particle-hole symmetry and the ν=5/2 quantum Hall state. *Phys Rev Lett* 2007; **99**: 236807.
71. Levin, M, Halperin, BI, Rosenow, B. Particle-hole symmetry and the Pfaffian state. *Phys Rev Lett* 2007; **99**: 236806.
72. Rezayi, EH, Haldane, FDM. Incompressible paired Hall state, stripe order, and the composite fermion liquid phase in half-filled Landau levels. *Phys Rev Lett* 2000; **84**: 4685.
73. Wan, X, Yang, K, Rezayi, EH. Edge excitations and non-Abelian statistics in the Moore-Read state: a numerical study in the presence of coulomb interaction and edge confinement. *Phys Rev Lett* 2006; **97**: 256804.
74. Möller, G, Simon, SH. Paired composite-fermion wave functions. *Phys Rev B* 2008; **77**: 075319.
75. Peterson, MR, Jolicoeur, T, Das Sarma, S. Finite-layer thickness stabilizes the Pfaffian state for the 5/2 fractional quantum Hall effect: wave function overlap and topological degeneracy. *Phys Rev Lett* 2008; **101**: 016807.

76. Peterson, MR, Park, K, Das Sarma, S. Spontaneous particle-hole symmetry breaking in the ν=5/2 fractional quantum Hall effect. *Phys Rev Lett* 2008; **101**: 156803.
77. Feiguin, AE, Rezayi, E, Yang, K, *et al.* Spin polarization of the ν=5/2 quantum Hall state. *Phys Rev B* 2009; **79**: 115322.
78. Wang, H, Sheng, DN, Haldane, FDM. Particle-hole symmetry breaking and the ν=5/2 fractional quantum Hall effect. *Phys Rev B* 2009; **80**: 241311.
79. Storni, M, Morf, RH, Das Sarma, S. Fractional Quantum Hall State at ν=5/2. *Phys Rev Lett* 2010; **104**: 076803.
80. Wójs, A, Tőke, C, Jain, JK. Landau-level mixing and the emergence of Pfaffian excitations for the 5/2 fractional quantum Hall effect. *Phys Rev Lett* 2010; **105**: 096802.
81. Möller, G, Wójs, A, Cooper, NR. Neutral fermion excitations in the Moore-Read state at filling factor ν=5/2. *Phys Rev Lett* 2011; **107**: 036803.
82. Rezayi, EH, Simon, SH. Breaking of particle-hole symmetry by Landau level mixing in the ν=5/2 quantized Hall state. *Phys Rev Lett* 2011; **106**: 116801.
83. Storni, M, Morf, RH. Localized quasiholes and the Majorana fermion in fractional quantum Hall state at ν=5/2 via direct diagonalization. *Phys Rev B* 2011; **83**: 195306.
84. Biddle, J, Peterson, MR, Das Sarma, S. Variational monte carlo study of spin-polarization stability of fractional quantum Hall states against realistic effects in half-filled landau levels. *Phys Rev B* 2013; **87**: 235134.
85. Tőke, C, Jain, JK. Understanding the 5/2 fractional quantum Hall effect without the Pfaffian wave function. *Phys Rev Lett* 2006; **96**: 246805.
86. Wójs, A, Quinn, JJ. Landau level mixing in the ν=5/2 fractional quantum Hall state. *Phys Rev B* 2006; **74**: 235319.
87. Tőke, C, Regnault, N, Jain, JK. Nature of excitations of the 5/2 fractional quantum Hall effect. *Phys Rev Lett* 2007; **98**: 036806.
88. Tőke, C, Regnault, N, Jain, JK. Numerical studies of the Pfaffian model of the fractional quantum Hall effect. *Solid State Commun* 2007; **144**: 504.
89. Dolev, M, Heiblum, M, Umansky, V, *et al.* Observation of a quarter of an electron charge at the ν=5/2 quantum Hall state. *Nature* 2008; **452**: 829.
90. Halperin, BI. Theory of the quantized Hall conductance. *Helv Phys Acta* 1983; **56**: 75.
91. Halperin, BI, Lee, PA, Read, N. Theory of the half-filled Landau level. *Phys Rev B* 1993; **47**: 7312.
92. Yang, G, Feldman, DE. Influence of device geometry on tunneling in the ν=5/2 quantum Hall liquid. *Phys Rev B* 2013; **88**: 085317.
93. Wen, XG, Niu, Q. Ground-state degeneracy of the fractional quantum Hall states in the presence of a random potential and on high-genus Riemann surfaces. *Phys Rev B* 1990; **41**: 9377.
94. Wen, XG. Non-Abelian statistics in the fractional quantum Hall states. *Phys Rev Lett* 1991; **66**: 802.
95. Wen, XG, Zee, A. Classification of Abelian quantum Hall states and matrix formulation of topological fluids. *Phys Rev B* 1992; **46**: 2290.
96. Yang, G, Feldman, DE. Experimental constraints and a possible quantum Hall state at ν=5/2. *arXiv: 1406.2263* 2014.
97. Samkharadze, N, Kumar, A, Manfra, MJ, *et al.* Integrated electronic transport and thermometry at milliKelvin temperatures and in strong magnetic fields. *Rev Sci Instrum* 2011; **82**: 053902.
98. Eisenstein, JP, Cooper, KB, Pfeiffer, LN, *et al.* Insulating and fractional quantum hall states in the first excited landau level. *Phys Rev Lett* 2002; **88**: 076801.
99. Xia, JS, Pan, W, Vicente, CL, *et al.* Electron correlation in the second Landau level: a competition between many nearly degenerate quantum phases. *Phys Rev Lett* 2004; **93**: 176809.
100. Pan, W, Xia, JS, Shvarts, V, *et al.* Exact quantization of the even-denominator fractional quantum Hall state at ν=5/2 landau level filling factor. *Phys Rev Lett* 1999; **83**: 3530.
101. Lilly, MP, Cooper, KB, Eisenstein, JP, *et al.* Evidence for an anisotropic state of two-dimensional electrons in high landau levels. *Phys Rev Lett* 1999; **82**: 394.
102. Du, RR, Tsui, DC, Stormer, HL, *et al.* Strongly anisotropic transport in higher two-dimensional Landau levels. *Solid State Commun* 1999; **109**: 389.
103. Pan, W, Du, RR, Stormer, HL, *et al.* Strongly anisotropic electronic transport at Landau level filling factor ν=9/2 and ν=5/2 under a tilted magnetic field. *Phys Rev Lett* 1999; **83**: 820.
104. Deng, N, Watson, JD, Rokhinson, LP, *et al.* Contrasting energy scales of reentrant integer quantum Hall states. *Phys Rev B* 2012; **86**: 201301(R).
105. Manfra, MJ, de Picciotto, R, Jiang, Z, *et al.* Impact of spin-orbit coupling on quantum Hall nematic phases. *Phys Rev Lett* 2007; **98**: 206804.
106. Koduvayur, SP, Lyanda-Geller, Y, Khlebnikov, S, *et al.* Effect of strain on stripe phases in the quantum Hall regime. *Phys Rev Lett* 2011; **106**: 016804.
107. Gervais, G, Engel, LW, Stormer, HL, *et al.* Competition between a fractional quantum Hall liquid and bubble and Wigner crystal phases in the third Landau level. *Phys Rev Lett* 2004; **93**: 266804.
108. Gamez, G, Muraki, K. ν=5/2 fractional quantum Hall state in low-mobility electron systems: Different roles of disorder. *Phys Rev B* 2013; **88**: 075308.
109. Pan, W, Masuhara, N, Sullivan, NS, *et al.* Impact of disorder on the 5/2 fractional quantum Hall state. *Phys Rev Lett* 2011; **106**: 206806.

110. Nuebler, J, Umansky, V, Morf, R, *et al.* Density dependence of the 5/2 energy gap: Experiment and theory. *Phys Rev B* 2010; **81**: 035316.
111. Reichl, C, Chen, J, Baer, S, *et al.* Increasing the ν=5/2 gap energy: an analysis of MBE growth parameters. *New Journal of Physics* 2014; **16**: 023014.
112. Liu, Y, Kamburov, D, Shayegan, M, *et al.* Anomalous robustness of the ν=5/2 fractional quantum Hall state near a sharp phase boundary. *Phys Rev Lett* 2011; **107**: 176805.
113. Choi, HC, Kang, W, Das Sarma, S, *et al.* Activation gaps of fractional quantum Hall effect in the second Landau level. *Phys Rev B* 2008; **77**: 081301.
114. Dean, CR, Piot, BA, Hayden, P, *et al.* Intrinsic gap of the ν=5/2 fractional quantum Hall state. *Phys Rev Lett* 2008; **100**: 146803.
115. Samkharadze, N, Watson, JD, Gardner, G, *et al.* Quantitative analysis of the disorder broadening and the intrinsic gap for the ν=5/2 fractional quantum Hall state. *Phys Rev B* 2011; **84**: 121305(R).
116. Deng, N, Gardner, GC, Mondal, S, *et al.* ν=5/2 fractional quantum Hall state in the presence of alloy disorder. *Phys Rev Lett* 2014; **112**: 116804.
117. Ilani, S, Martin, J, Teitelbaum, E, *et al.* The microscopic nature of localization in the quantum Hall effect. *Nature* 2004; **427**: 328.
118. Steele, GA, Ashoori, RC, Pfeiffer, LN, *et al.* Imaging transport resonances in the quantum Hall effect. *Phys Rev Lett* 2005; **95**: 136804.
119. Jura, MP, Topinka, MA, Urban, L, *et al.* Unexpected features of branched flow through high-mobility two-dimensional electron gases. *Nat Phys* 2007; **3**: 841.
120. Das Sarma, S, Hwang, EH. Mobility versus quality in two-dimensional semiconductor structures. *Phys Rev B* 2014; **90**: 035425.
121. Das Sarma, S, Hwang, EH. Short-range disorder effects on electronic transport in two-dimensional semiconductor structures. *Phys Rev B* 2014; **89**: 121413.
122. Eisenstein, JP, Willett, R, Stormer, HL, *et al.* Collapse of the even-denominator fractional quantum Hall effect in tilted fields. *Phys Rev Lett* 1988; **61**: 997.
123. Csáthy, GA, Xia, JS, Vicente, CL, *et al.* Tilt-Induced Localization and Delocalization in the Second Landau Level. *Phys Rev Lett* 2005; **94**: 146801.
124. Dean, CR, Piot, BA, Hayden, P, *et al.* Contrasting behavior of the 5/2 and 7/3 fractional quantum Hall effect in a tilted field. *Phys Rev Lett* 2008; **101**: 186806.
125. Xia, J, Cvicek, V, Eisenstein, JP, *et al.* Tilt-Induced Anisotropic to Isotropic Phase Transition at ν=5/2. *Phys Rev Lett* 2010; **105**: 176807.
126. Zhang, C, Knuuttila, T, Dai, Y, *et al.* ν=5/2 fractional quantum Hall effect at 10 T: implications for the Pfaffian state. *Phys Rev Lett* 2010; **104**: 166801.
127. Xia, J, Eisenstein, JP, Pfeiffer, LN, *et al.* Evidence for a fractionally quantized Hall state with anisotropic longitudinal transport. *Nat Phys* 2011; **7**: 845.
128. Liu, G, Zhang, C, Tsui, DC, *et al.* Enhancement of the ν=5/2 fractional quantum Hall state in a small in-plane magnetic field. *Phys Rev Lett* 2012; **108**: 196805.
129. Liu, Y, Hasdemir, S, Shayegan, M, *et al.* Evidence for a ν=5/2 fractional quantum Hall nematic state in parallel magnetic fields. *Phys Rev B* 2013; **88**: 035307.
130. Wójs, A, Möller, G, Simon, SH, *et al.* Skyrmions in the Moore-Read State at ν=5/2. *Phys Rev Lett* 2010; **104**: 086801.
131. Stern, M, Plochocka, P, Umansky, V, *et al.* Optical probing of the spin polarization of the ν=5/2 quantum Hall state. *Phys Rev Lett* 2010; **105**: 096801.
132. Rhone, TD, Yan, J, Gallais, Y, *et al.* Rapid collapse of spin waves in nonuniform phases of the second Landau level. *Phys Rev Lett* 2011; **106**: 196805.
133. Stern, M, Piot, BA, Vardi, Y, *et al.* NMR probing of the spin polarization of the ν=5/2 quantum Hall state. *Phys Rev Lett* 2012; **108**: 066810.
134. Tiemann, L, Gamez, G, Kumada, N, *et al.* Unraveling the spin polarization of the ν=5/2 fractional quantum Hall state. *Science* 2012; **335**: 828.
135. Wurstbauer, U, West, KW, Pfeiffer, LN, *et al.* Resonant Inelastic Light Scattering Investigation of Low-Lying Gapped Excitations in the Quantum Fluid at ν=5/2. *Phys Rev Lett* 2013; **110**: 026801.
136. Pan, W, Serafin, A, Xia, JS, *et al.* Competing quantum Hall phases in the second Landau level in the low-density limit. *Phys Rev B* 2014; **89**: 241302.
137. Das Sarma, S, Gervais, G, Zhou, X. Energy gap and spin polarization in the 5/2 fractional quantum Hall effect. *Phys Rev B* 2010; **82**: 115330.
138. Jain, JK. The 5/2 enigma in a spin? *Physics* 2010; **3**: 71.
139. Nayak, C. Fractional quantum hall effect: Full tilt. *Nat Phys* 2011; **7**: 836.
140. Willett, RL. The quantum Hall effect at 5/2 filling factor. *Rep Prog Phys* 2013; **76**: 076501.
141. Dolev, M, Gross, Y, Chung, YC, *et al.* Dependence of the tunneling quasiparticle charge determined via shot noise measurements on the tunneling barrier and energetics. *Phys Rev B* 2010; **81**: 161303.
142. Carrega, M, Ferraro, D, Braggio, A, *et al.* Anomalous Charge Tunneling in Fractional Quantum Hall Edge States at a Filling Factor ν=5/2. *Phys Rev Lett* 2011; **107**: 146404.

143. Bid, A, Ofek, N, Inoue, H, *et al.* Observation of neutral modes in the fractional quantum Hall regime. *Nature* 2010; **466**: 585.
144. Inoue, H, Grivnin, A, Ronen, Y, *et al.* Proliferation of neutral modes in fractional quantum Hall states. *arXiv: 1312.7553* 2013.
145. Milovanović, M, Read, N. Edge excitations of paired fractional quantum Hall states. *Phys Rev B* 1996; **53**: 13559.
146. Roddaro, S, Pellegrini, V, Beltram, F, *et al.* Nonlinear quasiparticle tunneling between fractional quantum Hall edges. *Phys Rev Lett* 2003; **90**: 046805.
147. Roddaro, S, Pellegrini, V, Beltram, F, *et al.* Interedge strong-to-weak scattering evolution at a constriction in the fractional quantum Hall regime. *Phys Rev Lett* 2004; **93**: 046801.
148. Roddaro, S, Pellegrini, V, Beltram, F, *et al.* Particle-hole symmetric Luttinger liquids in a quantum Hall circuit. *Phys Rev Lett* 2005; **95**: 156804.
149. Fendley, P, Fisher, MPA, Nayak, C. Edge states and tunneling of non-Abelian quasiparticles in the $\nu=5/2$ quantum Hall state and p+ip superconductors. *Phys Rev B* 2007; **75**: 045317.
150. Miller, JB, Radu, IP, Zumbuhl, DM, *et al.* Fractional quantum Hall effect in a quantum point contact at filling fraction 5/2. *Nat Phys* 2007; **3**: 561.
151. Bishara, W, Nayak, C. Edge states and interferometers in the Pfaffian and anti-Pfaffian states of the $\nu=5/2$ quantum Hall system. *Phys Rev B* 2008; **77**: 165302.
152. Feiguin, A, Fendley, P, Fisher, MPA, *et al.* Nonequilibrium transport through a point contact in the $\nu=5/2$ non-Abelian quantum Hall state. *Phys Rev Lett* 2008; **101**: 236801.
153. Fiete, GA, Bishara, W, Nayak, C. Multichannel Kondo models in non-Abelian quantum Hall droplets. *Phys Rev Lett* 2008; **101**: 176801.
154. Das, S, Rao, S, Sen, D. Effect of inter-edge Coulomb interactions on transport through a point contact in a $\nu=5/2$ quantum Hall state. *EPL* 2009; **86**: 37010.
155. Lin, PV, Camino, FE, Goldman, VJ. Superperiods in interference of e/3 Laughlin quasiparticles encircling filling 2/5 fractional quantum Hall island. *Phys Rev B* 2009; **80**: 235301.
156. McClure, DT, Zhang, Y, Rosenow, B, *et al.* Edge-state velocity and coherence in a quantum Hall Fabry-Pérot interferometer. *Phys Rev Lett* 2009; **103**: 206806.
157. Ofek, N, Bid, A, Heiblum, M, *et al.* Role of interactions in an electronic Fabry-Perot interferometer operating in the quantum Hall effect regime. *Proc Natl Acad Sci USA* 2010; **107**: 5276.
158. Zhang, Y. Waves, Particles, and Interactions in Reduced Dimensions. *Ph.D. Thesis*. Harvard University; 2009.
159. Fendley, P, Ludwig, AWW, Saleur, H. Exact nonequilibrium transport through point contacts in quantum wires and fractional quantum Hall devices. *Phys Rev B* 1995; **52**: 8934.
160. Yang, K. Field theoretical description of quantum Hall edge reconstruction. *Phys Rev Lett* 2003; **91**: 036802.
161. Baer, S, Rössler, C, de Wiljes, EC, *et al.* Interplay of fractional quantum Hall states and localization in quantum point contacts. *Phys Rev B* 2014; **89**: 085424.
162. Das Sarma, S, Freedman, M, Nayak, C. Topologically Protected Qubits from a Possible Non-Abelian Fractional Quantum Hall State. *Phys Rev Lett* 2005; **94**: 166802.
163. Bonderson, P, Kitaev, A, Shtengel, K. Detecting non-Abelian statistics in the $\nu=5/2$ fractional quantum Hall state. *Phys Rev Lett* 2006; **96**: 016803.
164. Emil, JB, Anders, K. 'One-dimensional' theory of the quantum Hall system. *Journal of Statistical Mechanics: Theory and Experiment* 2006: L04001.
165. Feldman, DE, Kitaev, A. Detecting non-Abelian statistics with an electronic Mach-Zehnder interferometer. *Phys Rev Lett* 2006; **97**: 186803.
166. Hou, CY, Chamon, C. "Wormhole" Geometry for Entrapping Topologically Protected Qubits in Non-Abelian Quantum Hall States and Probing Them with Voltage and Noise Measurements. *Phys Rev Lett* 2006; **97**: 146802.
167. Law, KT, Feldman, DE, Gefen, Y. Electronic Mach-Zehnder interferometer as a tool to probe fractional statistics. *Phys Rev B* 2006; **74**: 045319.
168. Stern, A, Halperin, B. Proposed experiments to probe the non-Abelian $\nu=5/2$ quantum Hall state. *Phys Rev Lett* 2006; **96**: 016802.
169. Feldman, DE, Gefen, Y, Kitaev, A, *et al.* Shot noise in an anyonic Mach-Zehnder interferometer. *Phys Rev B* 2007; **76**: 085333.
170. Ponomarenko, VV, Averin, DV. Mach-Zehnder interferometer in the fractional quantum Hall regime. *Phys Rev Lett* 2007; **99**: 066803.
171. Bonderson, P, Shtengel, K, Slingerland, JK. Interferometry of non-Abelian anyons. *Ann Phys* 2008; **323**: 2709.
172. Feldman, DE, Li, F. Charge-statistics separation and probing non-Abelian states. *Phys Rev B* 2008; **78**: 161304.
173. Rosenow, B, Halperin, BI, Simon, SH, *et al.* Bulk-edge coupling in the non-Abelian $\nu=5/2$ quantum Hall interferometer. *Phys Rev Lett* 2008; **100**: 226803.
174. Roulleau, P, Portier, F, Roche, P, *et al.* Noise dephasing in edge states of the integer quantum Hall regime. *Phys Rev Lett* 2008; **101**: 186803.
175. Bishara, W, Nayak, C. Odd-even crossover in a non-Abelian $\nu=5/2$ interferometer. *Phys Rev B* 2009; **80**: 155304.
176. Rosenow, B, Halperin, B, Simon, S, *et al.* Exact solution for bulk-edge coupling in the non-Abelian $\nu=5/2$ quantum Hall interferometer. *Phys Rev B* 2009; **80**: 155305.

177. Ponomarenko, VV, Averin, DV. Braiding of anyonic quasiparticles in charge transfer statistics of a symmetric fractional edge-state Mach-Zehnder interferometer. *Phys Rev B* 2010; **82**: 205411.
178. Stern, A, Rosenow, B, Ilan, R, *et al.* Interference, Coulomb blockade, and the identification of non-Abelian quantum Hall states. *Phys Rev B* 2010; **82**: 085321.
179. Wang, C, Feldman, DE. Identification of 331 quantum Hall states with Mach-Zehnder interferometry. *Phys Rev B* 2010; **82**: 165314.
180. Campagnano, G, Zilberberg, O, Gornyi, IV, *et al.* Hanbury Brown-Twiss interference of anyons. *Phys Rev Lett* 2012; **109**: 106802.
181. Rosenow, B, Simon, SH. Telegraph noise and the Fabry-Perot quantum Hall interferometer. *Phys Rev B* 2012; **85**: 201302.
182. Smits, O, Slingerland, JK, Simon, SH. Tunneling current through fractional quantum Hall interferometers. *Phys Rev B* 2014; **89**: 045308.
183. Willett, RL, Nayak, C, Shtengel, K, *et al.* Magnetic-field-tuned Aharonov-Bohm oscillations and evidence for non-Abelian anyons at ν=5/2. *Phys Rev Lett* 2013; **111**: 186401.
184. Wan, X, Hu, Z-X, Rezayi, EH, *et al.* Fractional quantum Hall effect at ν=5/2: Ground states, non-Abelian quasiholes, and edge modes in a microscopic model. *Phys Rev B* 2008; **77**: 165316.
185. Bishara, W, Bonderson, P, Nayak, C, *et al.* Interferometric signature of non-Abelian anyons. *Phys Rev B* 2009; **80**: 155303.
186. Chen, H, Hu, Z-X, Yang, K, *et al.* Quasiparticle tunneling in the Moore-Read fractional quantum Hall state. *Phys Rev B* 2009; **80**: 235305.
187. Zhang, Y, McClure, DT, Levenson-Falk, EM, *et al.* Distinct signatures for Coulomb blockade and Aharonov-Bohm interference in electronic Fabry-Perot interferometers. *Phys Rev B* 2009; **79**: 241304.
188. An, S, Jiang, P, Choi, H, *et al.* Braiding of Abelian and non-Abelian anyons in the fractional quantum Hall effect. *arXiv: 1112.3400* 2011.
189. Tsukazaki, A, Ohtomo, A, Kita, T, *et al.* Quantum Hall effect in polar oxide heterostructures. *Science* 2007; **315**: 1388.
190. Joseph, F, Denis, M, Yusuke, K, *et al.* Magnesium doping controlled density and mobility of two-dimensional electron gas in $Mg_x Zn_{1-x}O$/ZnO heterostructures. *Applied Physics Express* 2011; **4**: 091101.
191. Tsukazaki, A, Akasaka, S, Nakahara, K, *et al.* Observation of the fractional quantum Hall effect in an oxide. *Nat Mater* 2010; **9**: 889.
192. Maryenko, D, Falson, J, Kozuka, Y, *et al.* Temperature-Dependent Magnetotransport around ν=1/2 in ZnO Heterostructures. *Phys Rev Lett* 2012; **108**: 186803.
193. Nelson, SF, Ismail, K, Nocera, JJ, *et al.* Observation of the fractional quantum Hall effect in Si/SiGe heterostructures. *Appl Phys Lett* 1992; **61**: 64.
194. Lu, TM, Pan, W, Tsui, DC, *et al.* Fractional quantum Hall effect of two-dimensional electrons in high-mobility Si/SiGe field-effect transistors. *Phys Rev B* 2012; **85**: 121307.
195. Castro Neto, AH, Guinea, F, Peres, NMR, *et al.* The electronic properties of graphene. *Rev Mod Phys* 2009; **81**: 109.
196. Das Sarma, S, Adam, S, Hwang, EH, *et al.* Electronic transport in two-dimensional graphene. *Rev Mod Phys* 2011; **83**: 407.
197. Lebègue, S, Björkman, T, Klintenberg, M, *et al.* Two-Dimensional Materials from Data Filtering and Ab Initio Calculations. *Phys Rev X* 2013; **3**: 031002.
198. Novoselov, KS, Geim, AK, Morozov, SV, *et al.* Two-dimensional gas of massless Dirac fermions in graphene. *Nature* 2005; **438**: 197.
199. Zhang, Y, Tan, YW, Stormer, HL, *et al.* Experimental observation of the quantum Hall effect and Berry's phase in graphene. *Nature* 2005; **438**: 201.
200. Zhang, Y, Jiang, Z, Small, JP, *et al.* Landau-level splitting in graphene in high magnetic fields. *Phys Rev Lett* 2006; **96**: 136806.
201. Du, X, Skachko, I, Duerr, F, *et al.* Fractional quantum Hall effect and insulating phase of Dirac electrons in graphene. *Nature* 2009; **462**: 192.
202. Dean, CR, Young, AF, Cadden-Zimansky, P, *et al.* Multicomponent fractional quantum Hall effect in graphene. *Nat Phys* 2011; **7**: 693.
203. Novoselov, KS, Jiang, Z, Zhang, Y, *et al.* Room-temperature quantum Hall effect in graphene. *Science* 2007; **315**: 1379.
204. Özyilmaz, B, Jarillo-Herrero, P, Efetov, D, *et al.* Electronic Transport and Quantum Hall Effect in Bipolar Graphene p-n-p Junctions. *Phys Rev Lett* 2007; **99**: 166804.
205. Williams, JR, DiCarlo, L, Marcus, CM. Quantum Hall effect in a gate-controlled p-n junction of graphene. *Science* 2007; **317**: 638.
206. Amet, F, Williams, JR, Watanabe, K, *et al.* Selective Equilibration of Spin-Polarized Quantum Hall Edge States in Graphene. *Phys Rev Lett* 2014; **112**: 196601.
207. Bolotin, KI, Ghahari, F, Shulman, MD, *et al.* Observation of the fractional quantum Hall effect in graphene. *Nature* 2009; **462**: 196.
208. Ki, DK, Morpurgo, AF. High-quality multiterminal suspended graphene devices. *Nano Lett* 2013; **13**: 5165.

209. Bao, W, Miao, F, Chen, Z, *et al.* Controlled ripple texturing of suspended graphene and ultrathin graphite membranes. *Nat Nanotech* 2009; **4**: 562.
210. Xue, J, Sanchez-Yamagishi, J, Bulmash, D, *et al.* Scanning tunnelling microscopy and spectroscopy of ultra-flat graphene on hexagonal boron nitride. *Nat Mater* 2011; **10**: 282.
211. Dean, CR, Young, AF, MericI, *et al.* Boron nitride substrates for high-quality graphene electronics. *Nat Nanotech* 2010; **5**: 722.
212. Ghahari, F, Zhao, Y, Cadden-Zimansky, P, *et al.* Measurement of the ν=1/3 fractional quantum Hall energy gap in suspended graphene. *Phys Rev Lett* 2011; **106**: 046801.
213. Feldman, BE, Krauss, B, Smet, JH, *et al.* Unconventional Sequence of Fractional Quantum Hall States in Suspended Graphene. *Science* 2012; **337**: 1196.
214. Lee, DS, Skákalová, V, Weitz, RT, *et al.* Transconductance Fluctuations as a Probe for Interaction-Induced Quantum Hall States in Graphene. *Phys Rev Lett* 2012; **109**: 056602.
215. Feldman, BE, Levin, AJ, Krauss, B, *et al.* Fractional Quantum Hall Phase Transitions and Four-Flux States in Graphene. *Phys Rev Lett* 2013; **111**: 076802.
216. Roulleau, P, Portier, F, Roche, P, *et al.* Direct measurement of the coherence length of edge states in the integer quantum Hall regime. *Phys Rev Lett* 2008; **100**: 126802.
217. Granger, G, Eisenstein, JP, Reno, JL. Observation of chiral heat transport in the quantum Hall regime. *Phys Rev Lett* 2009; **102**: 086803.
218. Otsuka, T, Abe, E, Iye, Y, *et al.* Probing local electronic states in the quantum Hall regime with a side-coupled quantum dot. *Phys Rev B* 2010; **81**: 245302.
219. Venkatachalam, V, Hart, S, Pfeiffer, L, *et al.* Local thermometry of neutral modes on the quantum Hall edge. *Nat Phys* 2012; **8**: 676.
220. Kawamura, M, Kono, K, Hashimoto, Y, *et al.* Spatial gradient of dynamic nuclear spin polarization induced by breakdown of the quantum Hall effect. *Phys Rev B* 2011; **83**: 041305.
221. Gokmen, T, Padmanabhan, M, Shayegan, M. Transference of transport anisotropy to composite fermions. *Nat Phys* 2010; **6**: 621.
222. Padmanabhan, M, Gokmen, T, Shayegan, M. Ferromagnetic Fractional Quantum Hall States in a Valley-Degenerate Two-Dimensional Electron System. *Phys Rev Lett* 2010; **104**: 016805.
223. Kamburov, D, Shayegan, M, Pfeiffer, LN, *et al.* Commensurability oscillations of hole-flux composite fermions. *Phys Rev Lett* 2012; **109**: 236401.
224. Shabani, J, Gokmen, T, Chiu, YT, *et al.* Evidence for developing fractional quantum Hall states at even denominator 1/2 and 1/4 fillings in asymmetric wide quantum wells. *Phys Rev Lett* 2009; **103**: 256802.
225. Nuebler, J, Friess, B, Umansky, V, *et al.* Quantized ν=5/2 state in a two-subband quantum Hall system. *Phys Rev Lett* 2012; **108**: 046804.
226. Liu, Y, Hasdemir, S, Kamburov, D, *et al.* Even-denominator fractional quantum Hall effect at a Landau level crossing. *Phys Rev B* 2014; **89**: 165313.
227. Deng, N, Kumar, A, Manfra, MJ, *et al.* Collective nature of the reentrant integer quantum Hall states in the second Landau level. *Phys Rev Lett* 2012; **108**: 086803.
228. Chickering, WE, Eisenstein, JP, Pfeiffer, LN, *et al.* Thermopower of two-dimensional electrons at filling factors ν=3/2 and 5/2. *Phys Rev B* 2010; **81**: 245319.
229. Chickering, WE, Eisenstein, JP, Pfeiffer, LN, *et al.* Thermoelectric response of fractional quantized Hall and reentrant insulating states in the N=1 Landau level. *Phys Rev B* 2013; **87**: 075302.
230. Ki, DK, Fal'ko, VI, Abanin, DA, *et al.* Observation of even denominator fractional quantum Hall effect in suspended bilayer graphene. *Nano Lett* 2014; **14**: 2135.
231. Tian, J, Jiang, Y, Childres, I, *et al.* Quantum Hall effect in monolayer-bilayer graphene planar junctions. *Phys Rev B* 2013; **88**: 125410.
232. Dean, CR, Wang, L, Maher, P, *et al.* Hofstadter/'s butterfly and the fractal quantum Hall effect in moire superlattices. *Nature* 2013; **497**: 598.
233. Song, YJ, Otte, AF, Kuk, Y, *et al.* High-resolution tunnelling spectroscopy of a graphene quartet. *Nature* 2010; **467**: 185.